# Optimization of mammography with respect to anatomical noise

E. Fredenberg,[a] B. Svensson,[b] M. Danielsson,[a] B. Lazzari,[c] and B. Cederström[a]

[a]Department of Physics, Royal Institute of Technology (KTH), AlbaNova University Center, 106 91 Stockholm, Sweden;

[b]Sectra Mamea AB, Smidesvägen 5, 171 41 Solna, Sweden;

[c]Medical Physics Unit, General Hospital of Pistoia, Via Pertini, 708 - 51100 Pistoia, Italy

**ABSTRACT**

Beam quality optimization in mammography traditionally considers detection of a target obscured by quantum noise on a homogenous background. It can be argued that this scheme does not correspond well to the clinical imaging task because real mammographic images contain a complex superposition of anatomical structures, resulting in anatomical noise that may dominate over quantum noise. Using a newly developed spectral mammography system, we measured the correlation and magnitude of the anatomical noise in a set of mammograms. The results from these measurements were used as input to an observer-model optimization that included quantum noise as well as anatomical noise. We found that, within this framework, the detectability of tumors and microcalcifications behaved very differently with respect to beam quality and dose. The results for small microcalcifications were similar to what traditional optimization methods would yield, which is to be expected since quantum noise dominates over anatomical noise at high spatial frequencies. For larger tumors, however, low-frequency anatomical noise was the limiting factor. Because anatomical structure has similar energy dependence as tumor contrast, optimal x-ray energy was significantly higher and the useful energy region wider than traditional methods suggest. Measurements on a tissue phantom confirmed these theoretical results. Furthermore, since quantum noise constitutes only a small fraction of the noise, the dose could be reduced substantially without sacrificing tumor detectability. Exposure settings used clinically are therefore not necessarily optimal for this imaging task. The impact of these findings on the mammographic imaging task as a whole is, however, at this stage unclear.

## 1. INTRODUCTION

Beam-quality optimization of x-ray imaging systems is crucial to minimize the dose to the patient.[1,2] This is particularly so in mammography, where a large number of women are exposed to radiation through screening programs.[3–6] In traditional optimization of mammography, a target on a flat background is usually considered, and the ratio between target contrast and quantum noise is used as a figure of merit. Contrast and quantum noise are both reduced by a harder spectrum due to higher transmission at higher energies, and there exists an optimal energy at which the ratio is maximized at a constant dose to the breast.

Nevertheless, the dominant source of distraction for many imaging tasks in mammography is not quantum noise, but the variability of the anatomical background.[7,8] The so-called anatomical noise has similar energy dependence as the contrast, and when it is taken into account, we can expect a shift in the optimum towards higher energies.[9] In addition, when anatomical noise dominates, the dose may be reduced with little loss in image quality, because, contrary to quantum noise, the anatomical noise has the same dose dependence as the signal difference between target and background.

The purpose of the present study is to investigate the effect of anatomical noise on beam-quality optimization in mammography using an observer model. The methodology is similar to what we have previously employed for photon-counting spectral imaging,[10,11] but here we consider a conventional detector without energy resolution.

Electronic mail: fberg@mi.physics.kth.se

## 2. MATERIAL AND METHODS

### 2.1. Observer modeling

The noise equivalent number of quanta (NEQ) is an efficient and wide-spread metric for system optimization.[12,13] While the standard NEQ only takes quantum and detector noise into account, the dominant source of distraction for many imaging tasks in mammography is the variability of the anatomical background, which is fractal.[7,8] Richard and Siewerdsen[14,15] proposed the generalized NEQ (GNEQ) framework as a way to include the anatomical noise in optimization of imaging systems:

$$\text{GNEQ}(\omega) = \frac{\langle I \rangle^2 T^2(\omega)}{S_Q(\omega) + S_A(\omega)}, \tag{1}$$

where $\omega$ is the spatial frequency in the radial direction, $\langle I \rangle$ is the expected image signal, and $T$ is the modulation transfer function (MTF). $S_Q$ and $S_A$ are the power spectra (NPS) of quantum and anatomical noise respectively. For uncorrelated pixels, $S_Q$ is flat. $S_A$ can be described by a power law, $\alpha/\omega^\beta$, where $\alpha$ and $\beta$ are empirically determined constants, representing the magnitude and correlation of the noise, respectively.[7,8] Hence, anatomical and quantum noise have completely different frequency distributions; for large targets (e.g. tumors) the anatomical noise is of relatively large importance whereas for smaller objects (e.g. microcalcifications) the quantum noise dominates. It should be noted that the anatomical background of breast tissue in reality is not purely random, but has a deterministic component that reduces the magnitude of $S_A$.[8] This effect is not taken into account in the present study.

To take into account the frequency dependence and contrast of a specific task, we use an ideal-observer detectability index as a figure of merit:[13,14]

$$d'^2 = 2\pi \int_{\text{Ny}} \text{GNEQ}(\omega) \times C^2 \times F^2(\omega) \times \omega \, d\omega, \tag{2}$$

where the integral is taken over the Nyqvist region. $C = \Delta s / \langle I \rangle$ is the target-to-background contrast in the image for signal difference $\Delta s = \langle |I_{\text{background}} - I_{\text{target}}| \rangle$, where the angle brackets denote the expectation value. $F$ is the signal template, i.e. a spatial-frequency dependent task function, which integrates to the area of the target for unit contrast. Polar coordinates have been used for notational convenience, but all calculations were done in Cartesian coordinates where appropriate, i.e. rotational symmetry was not assumed in general. More advanced observer models, as well as inclusion of an eye filter and internal noise, have shown better correspondence with real observers in some cases,[15] but are not included in the present study.

In traditional optimization of mammography, where anatomical noise is disregarded, there exists an optimal energy for which the quotient $\Delta s^2 / S_Q$ and detectability is maximized. $S_A$, however, has similar energy dependence as $\Delta s^2$, and when it is taken into account, we can expect a flatter optimum and a shift towards higher energies, in particular for large objects. In addition, when $S_A$ dominates the dose may be reduced with no loss in image quality because $\Delta s^2 / S_A$ is dose independent as opposed to $\Delta s^2 / S_Q$. For a more formal derivation of noise and signal transfer in imaging systems we refer to our previous publications.[9–11]

### 2.2. Description of the system

The Sectra MicroDose Mammography system forms a basis for our study and it is briefly described here. The system consists of a tungsten-target x-ray tube with 0.5 mm aluminum filtration, a pre-collimator, and an image receptor, all mounted on a common arm (Fig. 1, Left). The image receptor consists of several modules of photon-counting silicon strip detectors with corresponding slits in the pre-collimator (Fig. 1, Right). To acquire an image, the arm is rotated around the center of the source so that the detector modules and pre-collimator are scanned across the object. The multi-slit geometry rejects virtually all scattered radiation.[16]

A bias voltage is applied over the silicon strip detector, so that when a photon interacts, charge is released and drifts as electron-hole pairs towards the anode and cathode respectively (Fig. 1, Right). Each strip is wire bonded to a preamplifier and shaper, which are fast enough to allow for single photon-counting. The preamplifier and shaper collect the charge and convert it to a pulse with a height that is proportional to the charge and thus

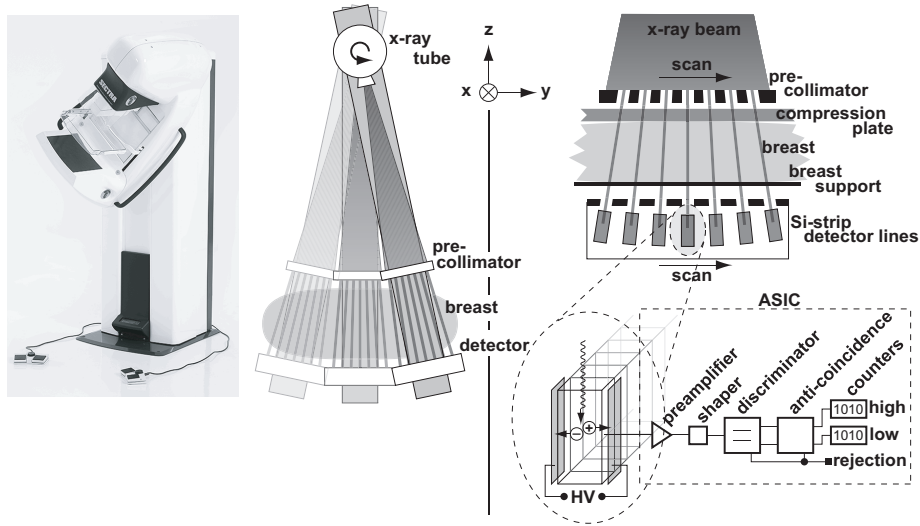

**Figure 1. Left:** Photograph and schematic of the Sectra MicroDose Mammography system [image courtesy Sectra Mamea AB]. **Right:** The image receptor and electronics.

to the energy of the impinging photon. Pulses below a few keV are regarded as noise and are rejected by a low-energy threshold in a discriminator. All remaining pulses are registered by counters. A preamplifier, shaper, and discriminator with counters are referred to as a channel, and all channels are implemented in an application specific integrated circuit (ASIC). Detailed descriptions of the system and detector can be found elsewhere.[5, 16]

As we have reported previously, an energy sensitive MicroDose system has been developed for spectral imaging within the EU-funded HighReX project.[9–11, 17, 18] The main difference compared to the conventional system is an additional high-energy threshold of the detector, which sorts the detected pulses into two bins according to energy.

We have designed a system model in the past that accurately predicts performance of both the conventional and spectral-imaging MicroDose systems.[9–11, 18] The model is based on detailed knowledge of the specific system geometry and published x-ray spectra,[19] attenuation coefficients,[20, 21] and average glandular dose (AGD) coefficients.[22] It was developed using the MATLAB software package (The MathWorks Inc., Natick, Massachusetts).

### 2.3. Measurement of the anatomical noise in mammograms

Measured values on the anatomical noise exponent, $\beta$, can be found in the literature and range between 1.5 and 4 for projection images of breast tissue ($\beta = 1.8 - 2.0$,[23] $\beta = 2.7$,[24] $\beta = 2.8$,[7] $\beta = 2.8 - 3.0$,[25] $\beta = 3$,[26] $\beta = 3.4 - 4.0$[27]), with mean value close to 3. It is clear that $\beta$ is affected by the system MTF, but, even if not corrected for, the MTF of most systems are close to unity at the low spatial frequencies where anatomical noise dominates. System parameters related to the x-ray spectrum, such as incident spectrum and detector efficiency, have no effect on the noise correlation at a first-order approximation.[11] Hence, $\beta$ can be regarded fairly system independent and the published values should be applicable to most mammography system.

Values on the anatomical noise magnitude, $\alpha$, are more difficult to deduce from published data, at least partly because $\alpha$ is to a larger degree system specific and does depend on, for instance, incident x-ray spectrum and detector efficiency. Nevertheless, knowledge of $\alpha$ is necessary to investigate the relative effect of quantum and anatomical noise, i.e. to find the crossing between the two. We have therefore measured the anatomical noise in a set of mammograms, acquired with a spectral MicroDose system at The Health Unit number 3 of Pistoia (Azienda USL3), Pistoia, Italy, as part of a clinical study within the HighReX project. A number of women from the screening program in the age group 50-69 were asked to participate. A set of images in CC view were used for the present study.

Assuming that the breast contains a mixture of adipose and glandular tissue within skin of constant thickness, contrast variation in a mammogram is caused by a variability in the glandular fraction superimposed on a thickness gradient. In a conventional non-energy-resolved mammogram there is no way of separating these two contrast mechanisms without using prior knowledge of, for instance, a slowly varying thickness gradient towards the breast border together with the mean glandularity. In addition, the thickness gradient interferes with low-frequency anatomical noise. Previous investigations have therefore been restricted to measurements in a central region where the compression paddle can be assumed to keep the breast at constant thickness.[26] The accuracy of this thickness measurement also limits the accuracy of the glandularity estimation, while the proposed method does not rely on a physical measurement of the breast thickness.

Using spectral information and the system model, however, we are able to separate thickness from glandularity. The model generates pixel values as a function of thickness and glandular fraction which are used to produce glandularity and thickness maps for the breast. For verification, the height of the compression paddle was registered for each image and compared to the 90% percentile of the thickness map.

Any other material in the breast that does not fit the breast model, such as skin at the breast border or microcalcifications, may generate outlier points. The continuous occurrence of such outlier points at the breast border was used as a starting point for the inner limit of the skin, given that it was at least 1 mm from the outer skin limit, which was determined by thresholding. Hence, a skin map could be removed from the glandularity map, and the remaining glandularity map was further eroded by a 2 mm-radius disc. Outlier points within the glandularity map were regarded as microcalcifications if the connected area was larger than one pixel. Isolated pixels were assumed to be statistical noise and were assigned values extrapolated from the neighboring pixels.

The glandularity maps were segmented into $256 \times 256$ pixel regions-of-interest (ROIs) with 128 pixels overlap. ROIs containing outlier points were discarded, and the image was excluded from measurement if more than 50% of the ROIs were discarded. The NPS was calculated in Cartesian coordinates as the mean of the squared Fourier transform of the difference in image signal from the mean in each ROI. The effect of the field of view on the measured NPS is a convolution with the window function (spectral leakage), and due to the sharply peaked spectrum, window artifacts can be expected, in particular at low spatial frequencies. A Hann data taper was therefore applied to each ROI prior to calculating the NPS. Rotational symmetry was assumed, and $S_A(\omega)$ was found by converting from Cartesian to polar coordinates and averaging over $2\pi$. The radial NPS was least-square fitted to

$$\ln(S) = \ln(\alpha \omega^{-\beta} + \Phi), \tag{3}$$

where $\Phi$ represents the quantum noise that is approximately flat. The logarithm makes the power-law NPS linear, which provides a better fit. Spatial frequencies above 0.1 mm$^{-1}$ were included in the fit. $\Phi$ was converted using the system model to an expected AGD, which was monitored to be within reasonable values in order to validate the fitting and conversion procedures.

Except for the exclusion criteria mentioned above, images were excluded from measurement if any of the quantities glandularity mean, $\alpha$, $\beta$, or quantum noise differed more than 50% between right and left breast image. Finally, a set of $2 \times 56$ images remained.

### 2.4. Optimization with respect to anatomical noise

#### 2.4.1. Computer simulation

As an example, we have chosen to illustrate the optimization procedure on the Sectra MicroDose Mammography system. It should be noted, however, that the results are generally valid for any mammography system. The anatomical noise levels were chosen in accordance with the measurement above, and the system model was used to evaluate the system response and to calculate the GNEQ and detectability index.

#### 2.4.2. Experimental verification

For the purpose of experimental verification, we have constructed a phantom simulating tissue with different glandularity content, as well as tumors with varying thickness embedded in tissue of 0, 50 and 100% glandularity, respectively. Since x-ray absorption to a good approximation can be described as a linear combination of only two functions, photo absorption and Compton scattering, which have separable energy and material dependencies, the absorption and energy dependence of any material can be simulated with a proper combination of any two materials.[21,28] In our case, the material bases are aluminum and polyethylene, and the phantom is made up of 25 squares with 30 mm side, where each cell contains a combination of aluminum and polyethylene representing a certain tissue type of 45 mm thickness. For example, 45 mm 50% glandular tissue is equivalent to 43.53 mm polyethylene and 1.04 mm Al, whereas the corresponding values for a 10 mm thick tumor embedded in 50% tissue are 43.65 mm and 1.25 mm, respectively.

The entrance surface air kerma (ESAK) for a fixed mAs was measured using an ion chamber (type 23344 and electrometer Unidose E, PTW, Freiburg, Germany) for tube voltages between 24 and 40 kV in steps of 1 kV. With knowledge of the total filtration, the AGD could be calculated, assuming a 45 mm breast with 50% glandularity.[22] Images were then acquired for a fixed AGD of 0.70 mGy. For each kV, two images were acquired with the phantom rotated 180° inbetween, in order to cancel any residual flat-fielding imperfections.

The mean pixel value and standard deviation was measured for each cell, using a ROI of $20 \times 20$ mm$^2$. Using these values, a scalar equivalent to Eq. 2 can be used

$$\text{SDNR}^2 \propto \frac{|m_i - m_j|^2}{(1-\chi)\sigma_{Q,i}^2 + \chi \sigma_A^2}, \qquad (4)$$

where $m_i$ and $\sigma_i$ is the mean and standard deviation, respectively, of the $i$:th ROI; $\chi$ is a parameter that shows the relative strength of the anatomical noise; and $\sigma_A$ is the standard deviation of the mean for an ensemble of ROIs with varying glandularity. Assuming that $T(\omega)$ has no energy-dependence, the dependence on beam quality of Eq. 2 calculated for a fixed dose, can be equally calculated using Eq. 4, if the noise mixing parameter, $\chi$, is chosen appropriately. We will make no attempt here to derive $\chi$, but only look at the two special cases $\chi = 0$ (statistical noise is completely dominant) and $\chi = 1$ (anatomical noise is completely dominant).

## 3. RESULTS AND DISCUSSION

### 3.1. Measurement of the anatomical noise in mammograms

The process of measuring the anatomical noise is shown for a representative case in Figure 2; conventional non-energy-resolved absorption image, thickness map, glandularity map, skin separation, and segmentation into ROIs. The radial NPS, measured in the glandularity map from Fig. 2 and fitted to Eq. 3, is shown in Fig. 3. In this example, the breast thickness was 5.7 cm and the mean glandularity was 0.29. NPS parameters $\alpha = 1.96 \times 10^{-4}$ mm$^2$ and $\beta = 2.75$, together with a dose of approximately 0.62 mGy yielded a crossing between anatomical and quantum noise at 3.4 mm$^{-1}$. Note that the markers in the NPS plot are interpolated from the actual measurement points to generate an equidistant log vector.

The results from the group of patients are summarized in Fig. 4, where (a) shows a histogram over the breast thickness that are approximately normal distributed around a mean of 5.0 cm. A scatter plot of the compression paddle height as a function of estimated thickness is shown in Fig. 4(b). There is a strong correlation but, irrespective of breast thickness, a systematic deviation of approximately 3 mm.

A scatter plot of the mean glandularities as a function of breast thickness is shown in Fig. 4(f). There seems to be an inverse correlation between thickness and glandularity, which is in accordance with previous studies.[29] Note, however, that the linear least-square fits that are displayed in Fig. 4(d–f) only serve to guide the eye; any correlations could not be established with confidence at this point and should only be regarded as indications. The mean glandularity over the group was 16%, which is surprisingly low compared to published results,[29] despite the relatively high age of the participating women (age group 50-69). The mean was calculated over the entire glandularity map and hence not biased by ROI sampling. Fig. 4(c) shows the mean glandularity distribution for the group, i.e. the average of the glandularity histograms calculated for each individual. We note that the glandularity distribution is skewed and not Gaussian.

Figures 4(d, e) show results from the NPS measurement; $\alpha$ and $\beta$ respectively. It seems $\beta$ is fairly invariant with breast thickness, whereas $\alpha$ exhibits anticorrelation. Note that, if the correlation observed here is a real effect, it has nothing to do with the passage of more material since the thickness is taken out of the equation. Rather, it may be explained by the fact that thicker breasts contain more adipose tissue (c.f. Fig. 4(f)), and

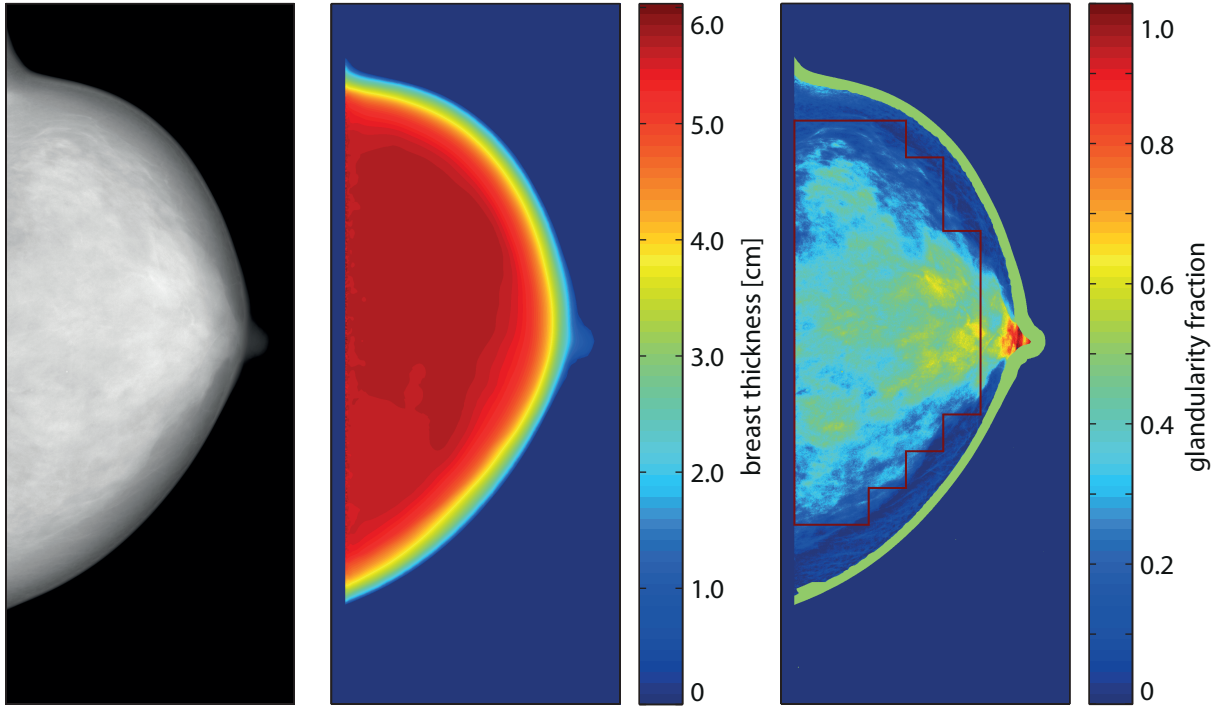

**Figure 2.** The process of deducing thickness and glandularity from a spectral mammogram. From left to right: 1) conventional absorption image where thickness gradient and glandular variation are superimposed, 2) thickness map, and 3) glandularity map, where the skin and ROI segmentations are highlighted. Color images available online.

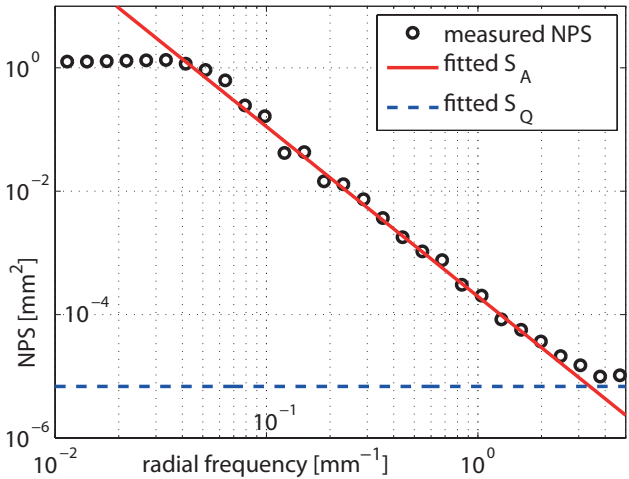

**Figure 3.** The radial NPS measured in the glandularity map of Fig. 2 and fitted to Eq. 3 to separate the anatomical ($S_A$) and quantum ($S_Q$) noise components.

hence less structure. When the line integral through the breast is taken into account in order to calculate the image NPS, these two effects will to some degree cancel. The mean $\alpha$ was $6.5 \times 10^{-5}$ mm$^2$, and the mean $\beta$ was 2.7 (standard deviation 0.06), which is in good agreement with previously reported values. These results can be used to calculate the NPS in breast images of virtually any imaging system, given that the system geometry is known.

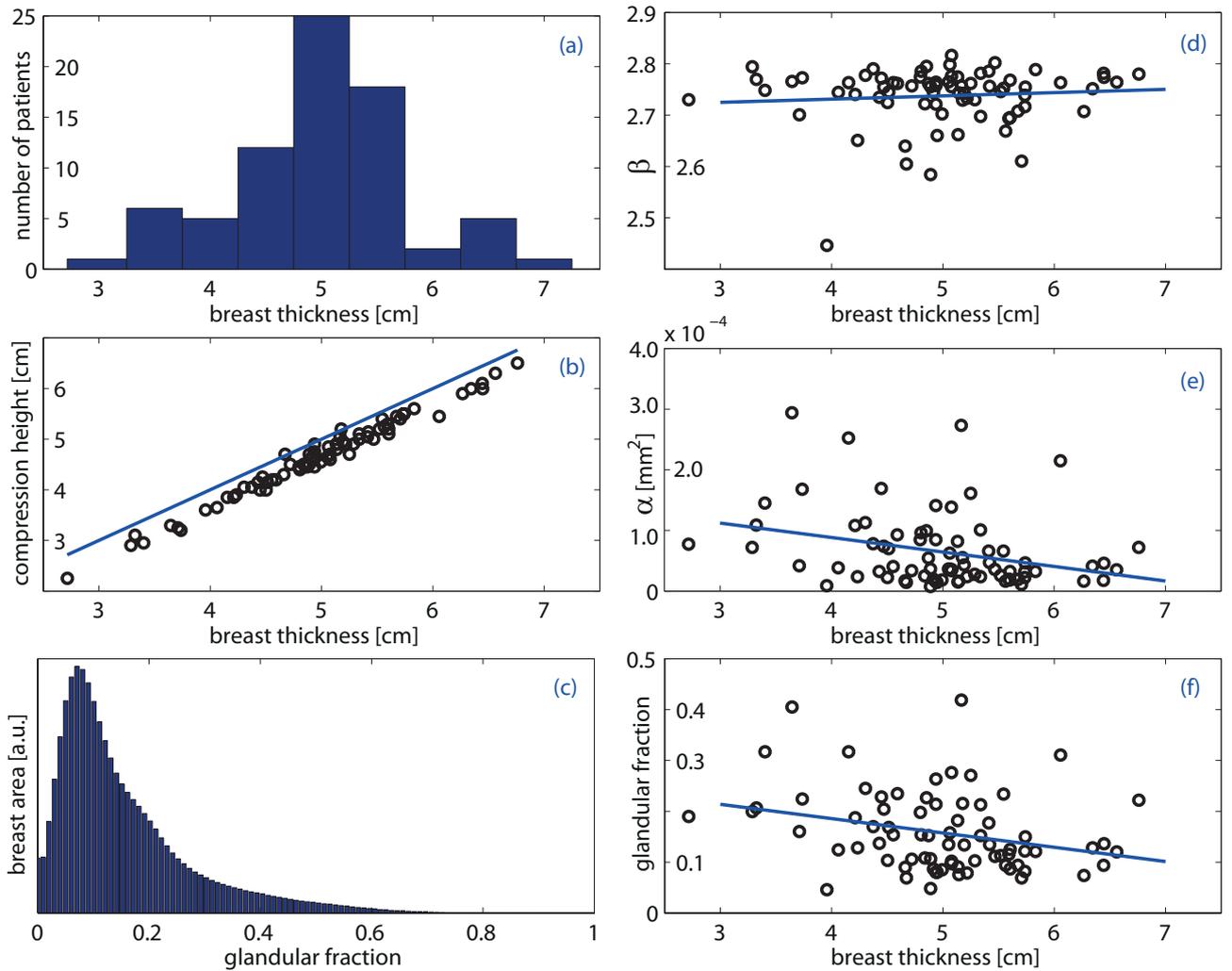

**Figure 4.** Measurement results from the set of mammograms: **(a)** histogram over breast thicknesses, **(b)** compression height as a function of estimated breast thickness, **(c)** mean glandularity distribution, **(d–f)** anatomical noise exponent ($\beta$), anatomical noise magnitude ($\alpha$), and glandularity, respectively, as functions of thickness. Measurement points are indicated by circles. The straight line in plot (b) shows the measured height in order to visualize the deviation. Linear least-square fits are provided in (d–f) in order to guide the eye to the trend.

### 3.2. Optimization with respect to anatomical noise

#### 3.2.1. Computer simulation

Figure 5 shows detectability ($d'^2$) as a function of x-ray tube acceleration voltage for a 5 mm tumor and a 100 $\mu$m microcalcification in quantum and anatomical noise. The targets were embedded in an average breast according to the previous section; 5.0 cm thickness, 20% glandularity, and a glandularity distribution according to Fig. 4(c). The mA was adjusted at each kV in order to keep AGD and exposure time constant. All three plots have equal axes and are directly comparable.

Figure 5(a) plots detectability for the tumor and microcalcification at an AGD of 0.5 mGy and in anatomical noise corresponding to $\alpha = 6.5 \times 10^{-5}$ mm$^2$ and $\beta = 2.7$, i.e. the mean values from the group of patients. The dashed lines represent detectability for pure anatomical noise and pure quantum noise. These extreme cases were normalized to the first data point for each target in order to show the trend over the kV range. The shaded areas represent the range where target detectability may fall depending on the relation between quantum and anatomical noise. We make the following observations:

- The pure-quantum-noise lines correspond to targets on flat backgrounds, i.e. similar to traditional optimization that only takes quantum noise into account. There is a maximum at 22 kV for both targets. Optima that are independent of target material is in line with previous research when quantum noise alone is considered.[6]

- The pure-anatomical-noise case has a less distinct maximum, which is shifted to a substantially higher kV compared to the quantum-noise case. Again, the optimum seems fairly independent of target material and is rather a function of target size.

- Detectability of the microcalcification is influenced by anatomical noise, but quantum noise dominates and the maximum is found at 24 kV, i.e. fairly close to traditional optimization.

- Tumor detectability is completely dominated by anatomical noise for this case, and the tube voltage dependence is significantly different from the outcome of traditional optimization methods.

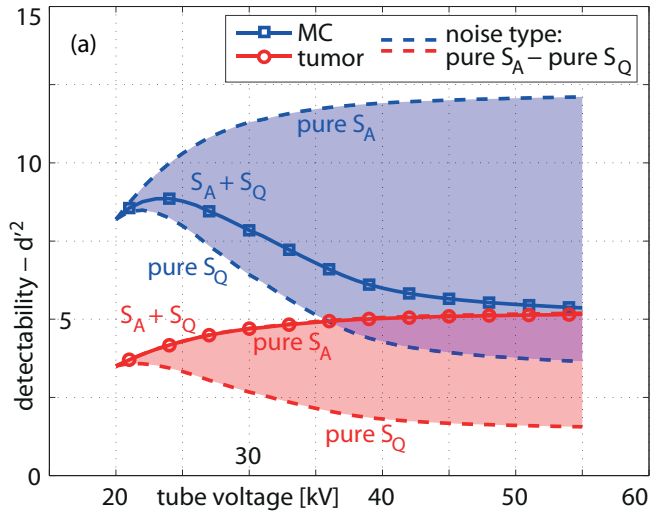

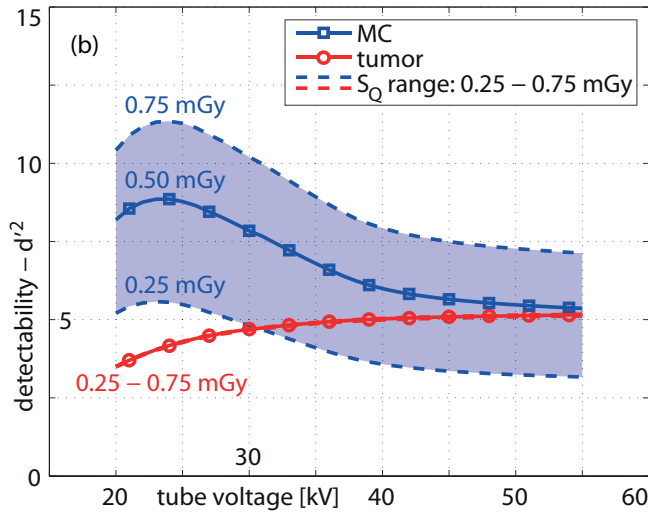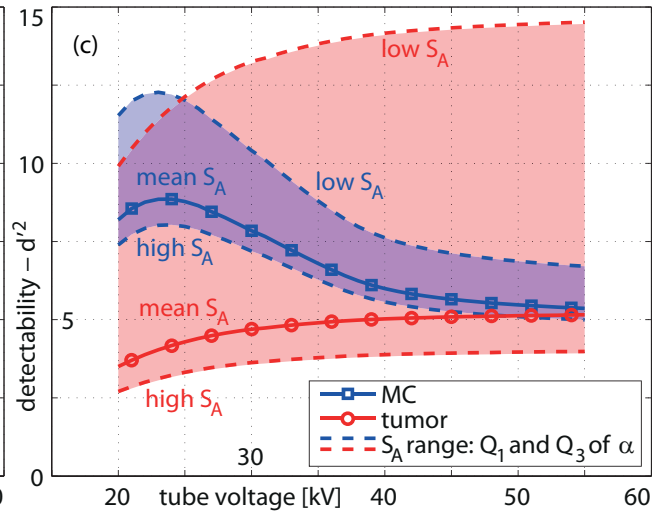

**Figure 5.** Detectability ($d'^2$) as a function of x-ray tube acceleration voltage for a 5 mm tumor and a 100 μm microcalcification (MC) in quantum $S_Q$ and anatomical $S_A$ noise. **(a)** Tumor and microcalcification at an AGD of 0.5 mGy and in an average level of anatomical noise. The dashed lines correspond to pure anatomical and pure quantum noise, normalized to the first data point. For each imaging task, the position of the solid line in the shaded area indicate the relative influence of the two noise sources. **(b)** Quantum noise dependence. The AGD over a range of dose levels in the shaded area, which yields a varying influence of quantum noise. For the tumor, the full range falls on the solid line and is therefore not visible. **(c)** Dependence on the anatomical noise magnitude with $\alpha$ ranging in the shaded area from the first ($Q_1$) to the third ($Q_3$) quartile of the measured values. Color images available online.

Quantum noise dependence of the two targets is illustrated in Fig. 5(b); detectability is plotted for dose levels 0.25–0.75 mGy, where the solid line in the center corresponds to the case in Fig. 5(a). Tumor detectability is virtually unaffected by dose and the three cases coincide. Detectability of the microcalcification, however, is clearly affected. For pure quantum noise, $d'^2$ would be proportional to dose, but because of the anatomical noise component in the microcalcification case there is a slightly weaker dependence, in particular at higher doses where anatomical noise becomes more important.

Figure 5(c) illustrates anatomical noise dependence in the range $\alpha \in (2.3, 8.4) \times 10^{-5}$ mm$^2$, which corresponded to the first and third quartile of the measured $\alpha$. The solid line is for the mean $\alpha$ and equals the corresponding lines in Fig. 5(a, b). Microcalcification detectability is slightly affected by the anatomical noise level, and the optimum is shifted in the order of 1 kV. The dependence is, however, substantially higher for the tumor case because if its larger size.

### 3.2.2. Experimental verification

We chose a background of 50% glandular tissue and as target a 20 mm tumor embedded in said tissue. For the anatomical noise, we used the five ROIs with glandularity equal to {0.3, 0.4, 0.5, 0.6, 0.7}. In Fig. 6, the squared SDNR is plotted as a function of tube voltage for $\chi = 0$ and $\chi = 1$, respectively. The curves are arbitrarily normalized to be equal at the tube voltage for which the statistical noise-dominated curve has its maximum. For comparison, theoretical curves of detectability index based on Eq. 2 are included. Even though the agreement is not prefect, the theoretical results are corroborated by the phantom experiment. Although the imaging cases are different, these curves may be qualitatively compared to the dashed curves in Fig. 5(a).

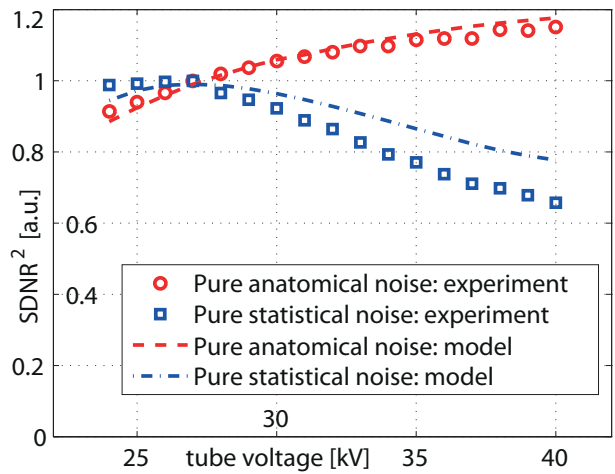

**Figure 6.** Squared signal-difference to noise ratio for a 20 mm tumor embedded in 50%-glandularity tissue as a function of tube voltage for pure statistical and pure anatomical noise, respectively. Also included are theoretical curves from Eq. 2.

## 4. CONCLUSIONS

Spectral information was used for advanced thickness equalization in a set of mammograms, and we were able to measure the anatomical noise without bias from a thickness gradient. Our detailed knowledge of the imaging system under study allowed us to convert the measured NPS in units of incident quanta to units of glandularity variations in the breast, which can be converted and used for virtually any mammography system. The measured anatomical NPS served as input to calculate an ideal-observer detectability index for, which was used to evaluate the system for a range of imaging conditions under influence of anatomical noise.

Figure 7 illustrates the main conclusions of this study; detection of a tumor and a microcalcification as a function of x-ray tube acceleration voltage. It is clear that inclusion of anatomical noise and imaging task in beam-quality optimization may yield very different results than a traditional analysis based solely on quantum noise. Detection of very small objects, such as small microcalcifications, is fairly unaffected by anatomical noise, and traditional optimization of the signal-to-quantum noise ratio is a reasonably good approximation. For larger targets, such as tumors, anatomical noise dominates over quantum, however, and optimum imaging conditions are shifted to considerably higher energies. This means that 1) it is possible to reduce dose without loss in image quality, and 2) it is also possible, or even beneficial, to increase the acceleration voltage. The latter may in turn be advantageous in order to reduce scan time or tube loading. These theoretical results were verified by measurements on a custom-made phantom that accurately simulates absorption of normal and cancerous tissue as a function of energy.

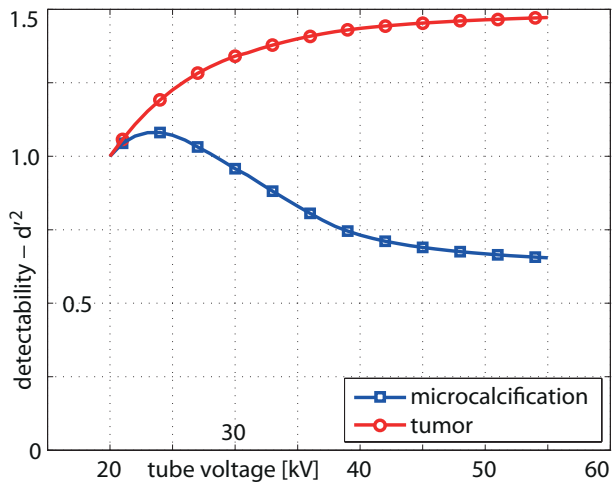

**Figure 7.** Detection of a tumor and a microcalcification as a function of x-ray tube acceleration voltage, i.e. the case in Fig. 5 but normalized to the first data point. The optima for these two targets are located at radically different voltages due to the inclusion of anatomical noise.

How these findings influence the optimization of the mammographic imaging task as a whole is at this stage unclear. It should also be noted that this is a simplified theoretical study and that these results should be corroborated by clinical investigations.

## ACKNOWLEDGMENTS

The authors wish to thank S. Karlsson and N. Dahlman at KTH, Stockholm, Sweden for phantom fabrication. This research was funded in part by the Swedish agency for innovation systems (VINNOVA).